\documentstyle[aps,prl]{revtex}

\newcommand{\be}{\begin{equation}}      
\newcommand{\ee}{\end{equation}}
\newcommand{\bea}{\begin{eqnarray}}     
\newcommand{\eea}{\end{eqnarray}}
\newcommand{\beb}{\begin{eqnarray*}}    
\newcommand{\eeb}{\end{eqnarray*}}
\renewcommand{\phi}{\varphi}

\newcommand{\UV}{{\sigma}}

\begin{document}



\twocolumn[\hsize\textwidth\columnwidth\hsize\csname
@twocolumnfalse\endcsname
\title{Dissipationless transport in low density bilayer systems}
\author{ Ady Stern$^a$, S. Das Sarma$^b$, Matthew P. A. Fisher$^c$ and
S. M. Girvin$^d$}
\address{
{\it (a)} Department of Condensed Matter Physics, Weizmann Institute,
Rehovot 76100, Israel\\
{\it (b)} Department of Physics, University of Maryland, College Park,
Maryland 20742-4111\\
{\it (c)} Institute for Theoretical Physics, UCSB, Santa Barbara,
California 93106-4030\\
{\it (d)} Department of Physics, Indiana University, Bloomington,
Indiana
47405-7105}

\date{\today}
\maketitle

\begin{abstract}
In a bilayer electronic system the layer index may be viewed as the
$z$-component of an isospin-$\frac{1}{2}$. An XY isospin-ordered
ferromagnetic phase
was observed in quantum Hall systems and is  predicted to exist at zero
magnetic field at low density. This phase is a superfluid for opposite
currents in the two layers.  At $B=0$ the system is gapless but
superfluidity is not destroyed by weak disorder.  In the quantum Hall
case,
weak disorder generates a random gauge field
which probably does not destroy              
superfluidity. Experimental signatures include Coulomb drag and
collective
mode measurements.
\end{abstract}
\pacs{73.40.Hm,73.20.Dx,75.10.Lp,75.10.Hk}
\vskip2pc]

\tighten
\narrowtext
In quantum well structures containing two separate
two-dimensional electron gases in close proximity,
an electron is described in terms of its position in the plane,
its spin and its layer index. The latter can be  regarded as an
isospin-$\frac{1}{2}$, denoted by ${\bf m}$,  with the two layers being
the
two eigenstates of $m_z$. States with spontaneous $XY$
isospin-ferromagnetic order have been
observed in quantum Hall systems\cite{jpebook} at total
Landau level filling factor $\nu=1$ and were  predicted to
exist at $B=0$ for sufficiently low electron
density.\cite{dassarma2} The origin of isospin ferromagnetism is a
favorable Coulomb exchange energy just as in the ordinary
Stoner instability.

For layer separation $d=0$ the isospin polarized phase breaks an
SU(2) symmetry, and the problem maps  onto the  Stoner
instability.\cite{jpebook}
For $d>0$, and in the absence of tunneling between the layers,  there
is an easy plane anisotropy since the direct Coulomb energy
favors polarization in the $XY$ plane ($\langle m_z\rangle=0$) in order
to avoid the cost of
charge imbalance between the layers that occurs for $\langle
m_z\rangle\ne 0$.
The angle of the magnetization ${\bf m}({\bf r})$
relative to the $x$--axis is then described by a
field $\phi({\bf r})$. Because the `charge' conjugate to the phase
$\phi$ is $m_z$, the Goldstone mode \cite{wenzee} associated with the
broken
U(1) symmetry at finite $d$ corresponds to superfluid currents which
are {\em opposite} in each layer. \cite{moon,jpebook,smgleshouches}

In this paper we study transport properties of the easy plane isospin
ferromagnet, focusing on the effects of disorder.   At $B=0$
we find that disorder weakens, but does not destroy the `gapless
isospin'
superfluidity.  The lack of time-reversal symmetry in the QHE case
causes disorder to induce a random gauge field which frustrates the
system, but
the Kosterlitz-Thouless transition probably survives weak disorder.
The effect of random interlayer tunneling is a separate and different
question. \cite{read95}

For $B=0$ it is
difficult to quantify the range of parameters, particularly $r_s$, in
which the isospin ferromagnetic state is the prevailing phase. 
It should lie between the low $r_s$ paramagnetic range and the
very high
$r_s$ range, where the system forms a
bilayer Wigner crystal. For a single layer, Monte Carlo calculations
\cite{r1} find a ferromagnetic transition
in the $r_s \sim 20-30$ range, and Wigner crystallization at
$r_s\approx
37\pm5$\cite{tanatar-ceperley}.
The energy differences among the various
possible phases are, however, very small \cite{r1} so that a definitive
statement is not possible -- in fact, earlier calculations
\cite{tanatar-ceperley} did not find a ferromagnetic transition. For a
double layer system,
Hartree-Fock (HF) theory predicts the existence of a broken symmetry isospin
ferromagnetic phase.\cite{dassarma2}  HF tends to
overestimate
the stability of broken
symmetry states, but its predictions are often
qualitatively correct and such states frequently do occur at
values of
$r_s$
larger than predicted. While quantum Monte
Carlo calculations are  needed to obtain the precise
 density at which a $B=0$ bilayer system will undergo the
spontaneous isospin ferromagnetic transition, it is reasonable to
assume,
based on  existing HF analysis \cite{dassarma2}, that such a
transition should occur at $r_s \approx 20-30$, a regime now
realizable in hole systems. \cite{yoonetal}

An HF analysis of the isospin polarized phase starts with an
Hubbard-Stratanovich decomposition of the Coulomb interaction, leading
to the action
\begin{eqnarray}
S&=&\int dt d{\bf r}\Big\{\psi^+i\partial_t \psi+
\frac{1}{2m}\psi^+[i{\bf
\nabla}-{\bf A}_aS^z]^2\psi({\bf r})\nonumber\\
&-& \rho_s({\bf r})V_H^{(1)}({\bf r}) - \rho^z({\bf r})V_H^{(2)}({\bf
r})\nonumber \\
&-& V_{ex}{\cal {\bf m}}\cdot \psi_\sigma^+({\bf r}){\bf
S}_{\sigma\sigma'}\psi_{\sigma'}({\bf r})\nonumber \\
&+&\frac{1}{2}V_{ex}{\bf m}^2+\frac{1}{2}\int d{\bf r'}n({\bf r})
V_s({\bf r}-{\bf r'})n({\bf r'})\nonumber\\
&+&\frac{1}{2}\int d{\bf r'}m_z({\bf r})V_a({\bf r} -{\bf r'})m_z({\bf
r'})
\Big\} \label{shf}
\end{eqnarray}
In Eq.~(\ref{shf}) $\psi^+,\psi$ are fermionic fields for the
electrons. The
symmetric and antisymmetric
densities are
$\rho_s\equiv \psi^+_\sigma\psi_\sigma$ and $\rho^z\equiv \psi^+_\sigma
S^z_{\sigma\sigma'}\psi_\sigma'$, with $S^i$ being the Pauli matrix for
the
$i$th-component of the isospin ($i=x,y,z$). The fields
$n({\bf r}),{\bf m}({\bf r})$ are auxiliary Hubbard-Stratanovich
fields
describing symmetric and
antisymmetric  densities. We are interested in the response of the
system to
a weak  antisymmetric vector potential ${\bf A}_a$, which is thus
included in the action (a factor of $\frac{e}{c}$ is
absorbed in ${\bf A}_a$). In momentum representation, $V_s({\bf
q})=\frac{2\pi e^2}{q}$ and $V_a({\bf q})=2\pi e^2 d$ (for small ${\bf
q}$).
For simplicity, we  assume here that the true
electron spin is fully aligned due to the Stoner instability
and can be ignored  (see however \cite{dassarma-sachdev}).

In momentum representation, for small ${\bf q}$, the symmetric Hartree
potential is
$V_H^{(1)}=V_s({\bf q})n({\bf q})$ while the antisymmetric is
$V_H^{(2)}=V_a({\bf q})
m_z({\bf q})$.
The Fock potential $V_{ex}$ is approximated in Eq.~(\ref{shf})
to be local, thereby neglecting the exchange contribution to the
gradient terms which contribute to  the isospin stiffness. We comment
on the actual value of $V_{ex}$ and on consequences of its non-zero
range below. In the  system's
response to ${\bf A}_a$ the symmetric field $n({\bf r})$ does not play
any role
and we omit it from following expressions.

 For fixed values of ${\bf m}$ and $n$,
the action (\ref{shf}) describes non-interacting electrons under the
influence of a space and time dependent scalar potential $V_H^{(1)}$,
vector potential ${\bf A}_a$ and Zeeman field $V_H^{(2)}{\hat
z}+V_{ex}{\bf m}$. In an $x-y$ ordered state, 
the saddle point for the bosonic fields is $n({\bf
r})=m_z({\bf r})=0$, and $|{\bf m}({\bf r})|={\cal M}$, a non-zero
constant.
Conventional approximation schemes (HF, random phase
approximation (RPA)) do not reliably obtain
${\cal M}$. Here  we
first assume full polarization (${\cal M}=n$, as predicted by
HF), and later discuss the case of partial polarization.

Due to an assumed lack of interlayer tunneling,
the action (\ref{shf}) possesses a $U(1)$ symmetry.\cite{wenzee}
 Thus, in
equilibrium the system picks an arbitrary direction for ${\bf m}$. We
write
${\bf m}= {\cal M}\left(\cos(\phi({\bf r})){\bf \hat x}+\sin(\phi({\bf
r})){\bf \hat y}\right)+m_z{\bf \hat z}$,
where $\phi$ is the angle between the
 planar component of the magnetization and the $\hat
x$--axis, and expect $\phi$ to be constant in the ground state and
slowly varying in low energy excitations. The energy cost of a
deviation from the equilibrium
magnetization is then expressed in terms of $\phi$ and $m_z$, and
should vanish for a uniform shift in
$\phi({\bf r})$.

We now integrate over the  fermionic fields and expand
the familiar ${\mbox{\rm tr}}\log\{\}$ term 
to second order in
$\phi$ and $m_z$. Within RPA the expansion is
given in terms of the response functions
$\chi_i\equiv-\langle\rho^i\rho^i\rangle$ and
$\chi_o\equiv-\langle\rho^z\rho^y\rangle$.
The effect of a slowly varying ${\bf A}_a$ on the $\chi$'s can be
separated out
by means of a Gorkov approximation, where ${\bf A}_a$ is approximated
not to
vary in the range of ${\bf r}-{\bf r'}$ and $t-t'$ in which the
response function is appreciable\cite{deGennes}.  The
effect of
${\bf A}_a$ is then incorporated by the ``minimal coupling'' prescription
$i{\bf\nabla}\phi\rightarrow (i{\bf \nabla}\phi-{\bf A}_a)$, and the
response
functions are calculated for ${\bf A}_a=0$.
The RPA action is then (omitting the zeroth order term),
\begin{eqnarray}
&&S_{RPA} \approx \frac{1}{2}\int
d\omega\int d{\bf r}  \Big\{\rho_s|i{\bf \nabla}{\phi}-
{\bf A}_a|^2 \nonumber\\ 
&+&\frac{e^2}{\Gamma}m_z^2
 +2\chi_o V_{ex}{\cal M}(V_{ex}+2\pi e^2 d)m_{z}{\phi} \Big\}
\label{seff2}
\end{eqnarray}
where
\begin{eqnarray}
 \rho_s&\equiv& -\lim_{{\bf q},\omega\rightarrow 0}
q^{-2}(1+\chi_yV_{ex})V_{ex}{\cal M}^2\\
 \frac{1}{\Gamma}&\equiv&-\lim_{{\bf q},\omega\rightarrow
0}[1+\chi_z(2\pi e^2d+V_{ex})](2\pi e^2d+V_{ex})
\label{seff2a}
\end{eqnarray}

The response functions $\chi_o,\chi_z,\chi_y$
are response functions of non-interacting
electrons in a
Zeeman field $V_{ex}{\cal M}{\hat x}$. For  small ${\bf q},\omega$,
\begin{eqnarray}
\chi_{z}&=&\chi_y=-\frac{1}{2}\Big[\frac{{\cal M}}{\Delta-\omega-{\cal
D}q^2}+
\frac{{\cal M}}{\Delta+\omega-{\cal D}q^2}\Big]\nonumber \\
\chi_{o}&=&-\frac{1}{2}\Big[\frac{i{\cal
M}}{\Delta-\omega-{\cal D}q^2}- \frac{i{\cal M}}{\Delta+\omega-{\cal
D}q^2}\Big]
\label{chis}
\end{eqnarray}
where $\Delta\equiv {\cal M}V_{ex}$ is the energy cost for flipping a
spin,
and the value of $\cal D$ is discussed below. The $U(1)$ invariance of
the problem is the reason for
$1+\chi_yV_{ex}$ being ${\cal O}(q^2)$.
The equation of
motion for $m_z$, derived from (\ref{seff2}),
is the Josephson type relation ${\dot\phi}=2\pi e^2d m_z$.

The integral over $m_z$ can now be carried out, resulting in an action
in terms of $\phi$ and ${\bf A}$ only, which is more transparent in
space and time representation:
\begin{equation}
\int dt\int d{\bf
r}\frac{{\dot\phi}^2}{2}[\frac{1}{2\pi
 e^2d}+\frac{1}{V_{ex}}]
+\frac{\cal MD}{2}|(i\nabla-{\bf A}_a)\phi|^2
\label{Sfinal}
\end{equation}
Eq.~(\ref{Sfinal}) is the action of a two-dimensional superfluid, with
$\cal MD$ being the
superfluid ``spin stiffness". 
{\em If ${\cal MD}>0$,
the bilayer system responds to 
the vector potential ${\bf A}_a$ as a superfluid. 
and an antisymmetric current flows without dissipation.}
Eq.  (\ref{Sfinal}) reveals the existence of a longitudinal Goldstone
mode
that carries antisymmetric density and satisfies
the dispersion relation
\begin{equation}
\omega^2={\cal MD}[\frac{1}{2\pi e^2d}+\frac{1}{V_{ex}}]^{-1}q^2.
\label{sound}
\end{equation}

Within RPA, the response functions $\chi_y,\chi_z$ are
\begin{eqnarray}
&&\chi_y({\bf q},\omega)=\chi_z({\bf q},\omega)
=\sum_{\alpha\beta} \frac{1}{2}|\langle\alpha|\rho_{\bf
q}|\beta\rangle|^2 \nonumber \\ 
&&\Big\{\frac{f(\epsilon_\alpha+\Delta)-
f(\epsilon_\beta)}{\omega+\Delta+\epsilon_\alpha -\epsilon_\beta+i\eta}
+\frac{f(\epsilon_\alpha)-f(\epsilon_\beta+\Delta)}{\omega
-\Delta+\epsilon_\alpha-\epsilon_\beta+i\eta}\Big\}
\label{generalpi}
\end{eqnarray}
where $|\alpha\rangle,|\beta\rangle$ are single particle
eigenstates of the spin-independent non-interacting Hamiltonian,
$\epsilon_\alpha,\epsilon_\beta$ are the corresponding single particle
energies,
$\rho_{\bf q}$ is the density operator and
$f(\epsilon)$ is the Fermi function.

Setting $\omega=0$ and expanding to second order in $q$,
we find that for a clean system with full isospin polarization
($2\Delta>\mu$,
$\mu$ being the chemical potential):
\begin{equation}
{\cal MD}=\int_{k<k_F} d{\bf k}
\left[\frac{1}{m}-\frac{k^2}{2m^2\Delta}\right]
=\frac{n}{m}(1-\frac{\mu}{2\Delta})
\label{sfd}
\end{equation}
This energy cost is the sum of single particle energies of eigenstates
$|{\bf k}\rangle$  of electrons in a Zeeman magnetic field that
precesses in space in a constant rate ${\bf\nabla}\phi$ and
is composed of one part
$\frac{({\bf\nabla}\phi)^2}{2m}$ originating from the anti-symmetric
current
induced by the precession of the field, and a second part,
$-\frac{({\bf k}\cdot{\bf\nabla}\phi)^2}{2m^2\Delta}$, which reflects
the slowing down of the symmetric motion due to the field precession.
There is no galilean invariance for antisymmetric
 currents so ${\cal MD}\ne n/m$.

The disorder potential can be
separated into symmetric and
antisymmetric parts. The symmetric part affects ${\cal D}$
much like non-magnetic disorder does in a conventional superconductor.
For weak symmetric disorder
($k_Fl\gg 1$, and hence $\Delta\tau\gg 1$), the disorder-averaged
matrix elements in
(\ref{generalpi}) are,
\begin{equation}
\left|\langle\alpha|\rho_{\bf
q}|\beta\rangle\right|^2=\frac{1}{\nu({\bar\epsilon})}\  
\frac{D({\bar\epsilon})q^2}
{(D({\bar\epsilon})q^2)^2+
(\epsilon_{\alpha}-\epsilon_\beta)^2}
\label{me}
\end{equation}
where
${\bar\epsilon}=\frac{1}{2}(\epsilon_\alpha+\epsilon_\beta)$,  $D$ is
the
diffusion constant, $\nu$ is the density of states, and
$|\epsilon_\alpha-\epsilon_\beta|<\frac{1}{\tau}$. Substituting in
(\ref{generalpi}) and paying attention to the dependences of $\nu$ and
$D$ on
$\bar\epsilon$ we find that the effect of symmetric disorder on the
spin stiffness (\ref{sfd}) is of order  $1/\Delta\tau$.

Antisymmetric  disorder
 modifies the capacitive
energy term in (\ref{shf}) to be
$
2\pi e^2 d\int d{\bf r}[m_z({\bf
r})-m_{z,{dis}}({\bf r})]^2
$
with random $m_{z,{dis}}$. As is known from  studies of, e.g.,
Josephson junction arrays,
such a randomization in the equilibrium distribution of $m_z$
reduces the superfluid density and can, if strong enough,
induce vortex--antivortex pairs destroying the superfluidity
even at zero temperature.  Here, since there are gapless Fermi surface
excitations even in the superfluid,
the resultant disordered phase may possibly be a normal
Fermi liquid with no long-range interlayer phase coherence.

Realistically, the disorder potential is made of comparable symmetric
and
antisymmetric components. For weak disorder,
then, antisymmetric currents flow without dissipation, although the
superfluid density is suppressed. As the disorder gets strong, it
eventually destroys the superfluidity.

The superfluid spin stiffness $\cal MD$ is also suppressed by finite
temperature.
Just as in an ordinary superconductor,    
its temperature dependence originates both from the Fermi functions in
(\ref{generalpi}) and from thermal fluctuations of vortex-antivortex
pairs in $\phi({\bf r})$. The spin stiffness, and with it long range
order
and antisymmetric
dissipationless transport, disppear entirely above a
Kosterlitz--Thouless (KT) transition temperature,
whose precise value depends on both effects.

An  experimental probe of  superfluidity of antisymmetric currents is
the transresistance, or drag
resistance, denoted by $\rho_{\mbox{\tiny D}}$. In a drag measurement
a current $I_1$ is driven
in one of the layers, while no current is allowed to pass through the
second
layer ($I_2=0$) which develops a voltage $V_2$. Then,
$\rho_{\mbox{\tiny
D}}\equiv -V_2/I_1$. For two identical layers, $\rho_{\mbox{\tiny D}}$
is
the difference between the symmetric and antisymmetric resistances. In
our case
the latter vanishes. Thus, {\it the transresistance equals the
symmetric one,
and the
voltages on the two layers should be equal in magnitude and direction}.
Since the superfluidity disappears at the K-T
transition temperature, $\rho_{\mbox{\tiny D}}$ would go {\em down}
with
increasing
temperature. Note that for weakly coupled
Fermi liquid bilayer
systems $\rho_{\mbox{\mbox{\tiny D}}}$ is opposite in sign to the
intra-layer
resistance,
and its magnitude {\em increases} 
 with temperature. If the superfluid mode is lost due
to disorder, the antisymmetric resistance becomes appreciable, and the
sign of
$\rho_{\mbox{\tiny D}}$ presumably becomes opposite to that of the
intra-layer
resistance.

The excitation of the sound mode
(\ref{sound})
is another  experimental probe.
In the absence of isospin ferromagnetism, a double layer system has
an antisymmetric acoustic plasmon mode, which  is overdamped by
disorder
as $q\rightarrow 0$.\cite{r3} Here, however, the sound mode
(\ref{sound})
is an underdamped Goldstone mode.  
A density sweep experiment through the
transition  will therefore exhibit a sharp  mode at low density which
will get overdamped
(at long wavelengths) above the transition density. Another distinction
between
these two collective modes is their
behavior when $d\rightarrow 0$.
In that limit the Goldstone mode will have a long wavelength
quadratic $q^2$ dispersion, since the $U(1)$ isospin symmetry changes
into an
$SU(2)$, whereas the normal acoustic plasmon mode tends toward
the single particle dispersion $v_Fq$.


So far we have taken the exchange Fock potential to be local, and
employed RPA. The $U(1)$ symmetry of the approximate actions
(\ref{seff2}) and (\ref{Sfinal}) is exact. However, other features of
our analysis are not, of which we expect two to be  most important.
First, as a consequence of its finite range, the Fock potential
renormalizes the dispersion relation of the electrons $\epsilon(k)$.
The sum (\ref{generalpi}) should then be evaluated with the
renormalized energy dispersion, leading to the replacements
$\frac{1}{m}\rightarrow \frac{\partial^2\epsilon}{\partial k^2}$ and
$\frac{k}{m}\rightarrow\frac{\partial\epsilon}{\partial k}$ in the
intergal in (\ref{sfd}), and affecting the spin stiffness.

Second, it is conceivable that the Stoner phase is only partially
isospin polarized, in contrast to the
HF prediction of full isospin polarization. Interestingly,
for a partially polarized state, and in the absence of disorder and
electron-electron interaction, the $q^2$-term in the sum
(\ref{generalpi}) vanishes (i.e., ${\cal D}=0$), due the constant
density of states $\nu$ characteristic of two dimensions. Spin
stiffness is then induced by the deviation of $\nu$ from a constant,
caused by the renormalization of the energy dispersion by interaction.
Similarly, in the presence of symmetric disorder, ${\cal D}\propto
[({D{\nu}})'{{\scriptstyle (\mu+\Delta/2)}}- ({D{\nu }})'{\scriptstyle
(\mu-\Delta/2)}] $, where a prime denotes differentiation with respect
to energy. Again, the energy dispersion must deviate from parabolic for
$\cal D$ to be non-zero.


The physics of the isospin ferromagnet at filling factor $\nu=1$ in the
QHE regime is quite different from that at $B=0$.
For two uncorrelated layers each
at filling factor $\nu=1/2$, there will be gapless `composite fermion'
excitations
In
the presence of interlayer phase coherence however, the finite isospin
stiffness leads to an energy gap for symmetric excitations and a QHE
plateau.
\cite{moon,jpebook,smgleshouches}

Because of the energy gap for symmetric excitations, the fermions can
be
reliably integrated out \cite{moon,jpebook,smgleshouches}
 to yield an Euclidean action which is a functional of the
isospin unit vector
${\hat{\UV}}({\bf r})\equiv{\bf m}({\bf r})/n({\bf r})$. The
Hamiltonian density corresponding to that action is
\begin{eqnarray}
H &=&
 \frac{1}{2} \bar\rho \partial_\mu \UV^\nu \partial_\mu \UV^\nu +
\frac{e^2}{2\Gamma} [n({\bf r}) \UV^z]^2\nonumber\\
&+& V_s({\bf r}) \delta n({\bf r}) + V_a({\bf r})[n_0 +
\delta n({\bf r})]\UV^z({\bf r}).
\label{eq:HAM}
\end{eqnarray}
$\Gamma$ is the double layer capacitance per unit area 
(including Hartree and exchange contributions), $V_{s,a}$ are the
symmetric and
antisymmetric parts of the disorder
potential
 and $\bar\rho \sim 1$K is the exchange-induced spin stiffness.
\cite{moon,smgleshouches}

A crucial consequence of the quantized Hall conductivity
 is a constraint relating the symmetric fermion
density $n({\bf r})=n_0+\delta n({\bf r})$ to the topological
(Pontryagin)
density
of the isospin field ${\bf\UV}$:
\cite{moon,jpebook,smgleshouches}
\be
\delta n({\bf r}) = \left(\frac{h}{e^2}\sigma_{xy}\right)\frac{1}{8\pi}
\epsilon^{\mu\nu}\epsilon_{abc}\, \UV^a\partial_\mu \UV^b\partial_\nu
\UV^c,
\ee
where Roman indices range over $x,y,z$ and Greek range over $x,y$, and
$n_0$ is
the
average density.

Taking advantage of the easy-plane anisotropy and noting that the XY
phase angle
field $\varphi$ contains vortex singularities, we
integrate out the massive $\sigma^z$ fluctuations
and find that the lack of time-reversal symmetry causes
the disorder potential to generate a gauge field yielding, in the high
temperature classical limit,
{\it a 2D XY model with random Dzyaloshinskii-Moriya
interaction }
\be
\int d{\bf r} \frac{\bar\rho }{2}|{\bf \nabla}\varphi + {\bf a}|^2
+ \sum_j  \left[\lambda V_a({\bf R}_j) - V_s({\bf R}_j)Q_j\right]M_j
\label{dmxy}
\ee
where $Q_j=\pm 1$ is the vorticity of the $j$th vortex (`meron'
\cite{moon,smgleshouches}),
$M_j=\pm 1$ is a flavor index indicating the sign of $\UV^z$ in the
vortex core, $\lambda$ is a non-universal constant related to the
core size, and we have dropped various irrelevant terms
(e.g., a random contribution to $\bar\rho$).  The gauge field is
$
{\bf a} \equiv \frac{\Gamma V_a}{4\pi\bar\rho e^2} {\bf J}_+,
$
where
${\bf J}_+= -\frac{h}{e^2}\sigma_{xy} \epsilon^{\mu\nu} \partial_\nu
V_s$ is
proportional to the symmetric Hall current. The $\Gamma V_a$ term is
the local
density imbalance.
If there is a Hall current flowing when there is a
density imbalance, then more current is flowing in one
layer than the other and the superfluid
mode \cite{moon,smgleshouches}
 ${{\bf J}_-}\sim \bar\rho({\bf \nabla}\varphi + {\bf a})$ is
necessarily
excited.
This is the physical interpretation of the gauge potential
${\bf a}$ which causes these currents to flow. The phase field $\phi$
in
(\ref{dmxy}) contains both the singular and smooth parts, and thus the
first
term in (\ref{dmxy}) mediates a logarithmic interaction between the
merons.
Unlike the $SU(2)$ symmetric case, where a symmetric disorder potential
introduces
a random gauge field \cite{Green}, in our case both symmetric and
anti-symmetric parts
are needed.


The gauge field
${\bf a}$ is random with a finite correlation length.
Ensemble averaging over a
closed contour
$\partial\gamma$ of perimeter $L$ gives
$
\big\langle \oint_{\partial\gamma}{\bf a}\cdot d{\bf r} \big\rangle =
0,
$
and
$
\big\langle \big[\oint_{\partial\gamma} {\bf a}\cdot d{\bf r}\big ]^2
\big\rangle \sim L^\theta
$
with $\theta=1$.  This is the gauge glass model for which it is known
that the KT transition is destroyed in the limit of strong disorder.
\cite{hyman}
For weak disorder
the phase diagram has proven difficult to determine, \cite{nattermann}
 but it is likely that the KT transition survives. 
This result can be motivated by noting that in order to have an isolated flux
quantum through a single plaquette, the vector potential would have
to fall off like $1/r$ and it follows
that the random potential $V_{s,a}$ would have to diverge.

To conclude, our results call for an experimental search for the $B=0$
phase coherent state in large $r_s$ bilayer
hole systems, and for a  study of how superfluidity in this phase and
its
analogous phase at $\nu=1$ is affected by disorder.   Superfluidity
could be
detected either in light scattering from the
Goldstone mode or by drag transport.

This work was supported by grants NSF DMR-97-04005 and DMR-95-28578
(MPAF), DMR-9714055 (SMG),
US-ONR (SDS), US-Israel BSF, Israeli Academy of Science, Victor Ehrlich
chair (AS)
and by NSF PHY94-07194 at the Institute for Theoretical Physics at
UCSB.

\end{document}